\documentstyle[12pt]{article}
\begin{document}
\renewcommand{\thesection}{\Roman{section}}
\baselineskip 20pt
{\hfill PUTP-95-23}                                  
\vskip 2cm
\begin{center}
{\large\bf Possible retardation effects of quark confinement \\
on the meson spectrum}
\end{center}
\vskip 7mm
\centerline{Cong-feng Qiao~~~ Han-Wen Huang~~~Kuang-Ta Chao}
\centerline{\small\it Center of Theoretical Physics, CCAST(World Laboratory), 
            Beijing 100080, P.R. China}
\centerline{\small\it Department of Physics, Peking University, 
            Beijing 100871, P.R. China}
\begin{center}
\begin{minipage}{120mm}
\vskip0.6in
\begin{center}{\bf Abstract}\end{center}
  { The reduced Bethe-Salpeter equation with scalar confinement and vector
gluon exchange is applied to quark-antiquark bound states.  
The so called intrinsic flaw of Salpeter equation with static 
scalar confinement is investigated. The notorious problem of narrow level
spacings is found to be remedied by taking into consideration the retardation 
effect of scalar confinement. Good fit for the mass spectrum of both heavy and 
light quarkomium states is then obtained.}
\vskip 0.5cm
PACS number(s):
\end{minipage}
\end{center}

\vfill\eject\pagestyle{plain}\setcounter{page}{1}

\section{Introduction}

{\hskip 7mm}To understand quark confinement is the most important task in 
{\bf QCD}
and hadron physics. Lattice {\bf QCD} calculations show that the interquark 
potential for a heavy quark-antiquark pair $Q\bar{Q}$ in the static limit is 
well described by a linear confining potential, plus a short-ranged Coulomb
potential\cite{la}. Phenomenologically, these potentials have been used with 
the Schr\"odinger equation for nonrelativistic heavy quarkonium systems like
$c\bar{c}$ and $b\bar{b}$ states, and satisfactory results for their mass 
spectrum have been obtained. Not only the spin-independent but also the 
spin-dependent $Q\bar{Q}$ potentials are studied both in lattice QCD\cite{lb}
and the quark potential model\cite{s1,s2,ding}. Most results seem to be consistent 
with the picture that the dominant part of linear confinement potential is 
transformed as a Lorentz scalar, while the Coulomb potential stems from one 
gluon exchange, which has the feature of a Lorentz vector. In particular, 
the fact that the spin-orbit term (Thomas procession term) induced by the 
scalar confining potential tends to partially compensate the spin-orbit term
generated by one gluon exchange, has been strongly supported by the observed
fine splittings of P-wave $c\bar{c}$ and $b\bar{b}$ states\cite{s1,s2,ding}.

However, a rather serious problem seems to remain if the spin-independent
relativistic correction, caused by the static (instantaneous) scalar confining
potential, is taken into consideration. This spin-independent term is of the 
same order as the Thomas precession term in the nonrelativistic expansion in 
terms of ${\vec p}^2/{m^2}$. As noted before\cite{ding}, including this 
spin-independent term 
\begin{equation}
\label{1}
H_{SI} = -\frac{1}{4 m^2} (2{\vec p}^2 S + 2S{\vec p}^2 + \frac{2}{r}
\frac{dS}{dr} + \frac{d^2 S}{d r^2}),
\end{equation}
where $S(r)$ is the static scalar confining potential and is usually assumed 
to
take the form of $S(r)=\lambda{r}$ with $\lambda$ being the string tension,
into the Hamiltonian will badly disturb the mass spectrum of mesons
(even for $c\bar{c}$
states), because this term is negative and unreasonably large for higher 
excited
states, making the level spacings for higher lying states unreasonably small. 
This problem is probably due to the fact that the scalar confining potential 
has been treated as an instantaneous potential, which is valid in the static 
limit but may not hold when relativistic corrections are taken into account.
This problem was also noted by other authors\cite{s3} in the framework of the
reduced Salpeter equation. This equation is equivalent to the Breit
equation to the first order of $\vec{p}^2/m^2$, and may be used to study
higher order relativistic corrections for systems containing the charm quark 
and even lighter quarks. It
was found \cite{s3} that in the framework of the reduced Salpeter 
equation with
an instantaneous scalar confining potential, the level spacings (e.g.,the $2S - 
1S $ spacing) would tend to vanish for $q\bar{q}$ mesons when the constituent 
quark mass approachs to zero, and difficulties are already evident for the
$c\bar{c}$ states. It was then pointed out\cite{s3} that there is an intrinsic
flaw in the approach which uses the reduced Salpeter equation with static 
scalar confinement potential.
%%%%%%%%%%%%%%%%

To overcome this difficulty, several scenarios have been put 
forward\cite{s3}\cite{s4}.
The chief differences between those works are of the useage of interaction 
potentials. By now, there are no mature theories or calculations for $ q\bar q $
confinement interaction in QCD, and customarily used potentials are phenomenological 
and have some uncertain perameters in them. For different procedure of 
evaluating 
these parameters, one can have some different ways to fit the experimental data. 
Therefore, to classify some of the most effective alternatives of 
quark-antiquark interaction potentials seems still premature. 

Despite of the limited understanding for confinement at present, more 
theoretical efforts should be made to study this problem. In our opinion, the
difficulty with the reduced Salpeter equation and the static scalar 
confinement  is probably due to the improper treatment that the confining 
interactiton is purely instantaneous. With some retardation effect of quark 
confinement being considered, even within the framework of reduced Salpeter
equation the level spacings for $q\bar{q}$ mesons could become normal since
the retardation effect might cancel the sick disturbance caused by the 
spin-independent correction from the instantaneous part of scalar confinement 
\cite{ch}. To implement this idea, we will assume the confinement kernel
in momentum space to take the form
\begin{equation}
\label{2}
G(q)\propto \frac{1}{(-q^2)^2}=\frac{1}{(\vec{q}^2 - q_0^2)^2},
\end{equation}
where $q$ is the $4$-momentum exchanged between the quark and antiquark in a
meson. In fact, this form was suggested for the dressed gluon propagator at
small momenta to implement confinement\cite{pa}. Here we will use the same 
form  but regard it as an effective scalar confinement kernel. Then, if the 
system is not highly relativistic we may make the approximation
\begin{equation}
\label{3} 
G(q)\propto\frac{1}{(\vec{q}^2 - q_0^2)^2}\approx\frac{1}{(\vec{q}^2)^2}
\left(1 + \frac{2 q_0^2}{\vec{q}^2}\right),
\end{equation}
and may further express $q_0$ in terms of its on-shell values which are 
obtained by assuming that quarks are on their mass shells. This should be a
good approximation for $c\bar{c}$ and  $b\bar{b}$ states, because they are 
nonrelativistic systems and the binding energies are smaller than the quark 
masses, therefore the quarks are nearly on their mass shells. In order to get a
qualitative feeling about the retardation effect considered here, 
we will also
use(\ref{3}) for light quark mesons, though the approximations are not as 
good as for heavy quark mesons. With above approximations, the scalar 
confinement kernel becomes instantaneous again but we have incorperated
some retardation effect into the kernel. In the static limit, the retardation
term vanishes and the kernel returns to $G(q)\propto \frac{1}{(\vec{q}^2)^2}$,
which is just the Fourier transform of the linear confining potential.

In this paper, we will use this
modified scalar confining potential in which the 
retardation effect is incorperated and
the one-gluon-exchange potential in the framework of the reduced Salpeter 
equation to study the mass spectra of $q\bar{q}$ mesons including both heavy 
and light mesons. We will concentrate on the $0^- $ and $1^-$ mesons to 
examine their level spacings.

\section{Reduced Salpeter equation with scalar and vector interactions}

{\hskip 7mm}In quantum field theory, a basic description for the bound states is the
Bethe-Salpeter equation\cite{s5}. Define the Bethe-Salpeter 
wave function of the 
bonud stase $ \mid P\rangle $  of a 
quark $ \psi(x_1) $ and an antiquark $ \bar{\psi}(x_2) $ as

\begin{equation}             
\label{e1}
\chi(x_1,x_2)=\langle 0\mid T\psi(x_1)\bar{\psi}(x_2)\mid P\rangle. 
\end{equation}
where $ T $ represents time-order product, and transform it into the momentum space
\begin{equation}
\label{e2}
\chi_P(q)=e^{{-iP}\cdot X}\int d^{4}x e^{{-iq}\cdot x} \chi(x_1,x_2).         
\end{equation}
Here $ P $ is the four-momentum of the meson 
and $ q $ is the relative momentum of quark and           
antiquark. 
We use the standard center of mass and relative variables:
\begin{equation}
X = \eta_1 x_1 + \eta_2 x_2,~~~x=x_1 - x_2,
\end{equation}
where $\eta_i=\frac{m_i}{m_1+m_2}~(i=1,2)$. Then in momentum space the bound
state BS equation reads
\begin{equation}
\label{e3}
(\not\!{p_1}-m_1)\chi_P(q)(\not\!{p}_2+m_2)=\frac{i}{2\pi}
\int d^{4}k G(P,q-k)\chi_P(k), 
\end{equation}
where $ p_1 $ and $ p_2 $ represent the momentum of 
quark and antiquark respectively.

\begin{equation}
\label{e4}
p_1=\eta_1 P+q ,~~~  p_2=\eta_2 P-q. 
\end{equation}
$ G(P,q-k) $ is the 
interaction kernel which acts on $ \chi $ and is determined by the 
interquark dynamics.
Note in Eq.(\ref{e3}) $m_1$ and $m_2$ represent the effective constituent 
quark masses so that we could use the effective free propagators of quarks 
instead of the full propogators. This is an improtant approximation and 
simplification for light quarks. Furthermore, because of the lack of a 
fundamental description for the nonperturbative QCD dynamics, 
we have to
make some approximations for the interaction kernel of quarks.
In solving Eq.(\ref{e3}), we 
assume the kernel to be instantaneous (but with some retardation effect in a
modified form for the kernel) and 
neglect the negative energy projectors 
in the quark propagators,  because in general the negative 
energy projectors  only contribute to quantities
of higher orders  due to $M - E_1 - E_2 \ll M + E_1 + E_2 $, where M, $E_1$, 
and $E_2$ are the meson mass, the quark kinetic energy, and the antiquark 
kinetic energy respectively.
Based on the above assumptions  the BS equation 
can be reduced to a three-dimensional equation, i.e., 
the reduced Salpeter equation, for the three dimensional BS wave function

\begin{equation}
\label{e10}
\Phi_{\vec P}(\vec q)=\int dq^0 \chi_P(q^0,\vec q),
\end{equation}

\begin{equation}
\label{e11}
(P^0-E_1-E_2)\Phi_{\vec P}(\vec q)=
\Lambda_{+}^{1}\gamma^0\int d^3 k G(\vec P,\vec q ,\vec k)\Phi_{\vec P}(\vec k)
\gamma^0 \Lambda_{-}^{2}.
\end{equation}
Here

$$ \Lambda_{+}^{1}={1\over 2E_1}(E_1+\gamma^0 {\vec {\gamma}}
    \cdot{\vec p_1}+m_1 \gamma^0 ), $$
\begin{equation}
\label{e12}
\Lambda_{-}^{2}={1\over 2E_2}(E_2-\gamma^0 {\vec {\gamma}}
\cdot{\vec p_2}-m_2 \gamma^0 ),
\end{equation}
are the remaining positive energy projectors of the quark and antiquark
respectively, and $E_1 = \sqrt{m_1^2+\vec{p_1}^2},~E_2 = 
\sqrt{m_2^2+\vec{p_2}^2}$. 
The formal products of $ G \Phi $ in Eq.(\ref{e11}) 
take the form

\begin{equation}
\label{e13}
G\Phi=\sum\limits_{i} G_i O_i \Phi O_i = G_s \Phi + \gamma_{\mu}\otimes
   \gamma^{\mu} G_v \Phi, 
\end{equation}
where $ O=\gamma_{\mu} $ corresponding to the pertubative 
one-gluon-exchange interaction and $ O=1 $ for the scalar 
confinement potential.
  % A space line  is necessary here
  
From Eq.(\ref{e11}) it is easy to see that
$$ \Lambda_{+}^{1}\Phi_{\vec P}(\vec q)=\Phi_{\vec P}(\vec q), $$
\begin{equation}
\label{e14}
\Phi_{\vec P}(\vec q)\Lambda_{-}^{2}=\Phi_{\vec P}(\vec q).
\end{equation}
Considering the constraint of Eq.(\ref{e14}), and the requirement 
of space reflection, 
in the rest frame of the meson 
$({\stackrel{\rightharpoonup }{P}}=0) $ the wave function 
$\Phi_{\vec P}(\vec q) $ for the $0^-$ and $1^-$ mesons can be written as 

\begin{eqnarray}
\label{a31}
&&\Phi _{\stackrel{\rightharpoonup }{P}}^{0^{-}}( \stackrel{%
\rightharpoonup }{q}) =\Lambda _{+}^1\gamma ^0( 1+\gamma ^0
) \gamma _5\gamma ^0\Lambda _{-
}^2\varphi(\stackrel{\rightharpoonup }{q}),\nonumber \\ 
&&\Phi _{\stackrel{\rightharpoonup }{P}}^{1^{-}}( \stackrel{%
\rightharpoonup }{q}) =\Lambda _{+}^1\gamma ^0( 1+\gamma ^0
) \not\!e\gamma ^0\Lambda _{-}^2f(\stackrel{\rightharpoonup }{q}),
\end{eqnarray}
where $\not\!e =\gamma_{\mu} e^\mu$, $e^\mu$ is the polarization vector of   
$1^-$ meson, and $\varphi(\vec{q})$, $f(\vec{q})$ are scalar functions
of $\vec{q}^2$.
It is easy to show that Eq.(\ref{a31}) is the most 
genernal form for the 
$0^-$ and $1^-$ (S-wave) $q_1 \bar{q_2}$ meson wave 
functions at the rest
frame [e.g., for the $0^-$ meson wave function there are four independent
scalar functions but with the constraint of 
Eq.(\ref{e14}) those scalar functions can be reduced 
to one and expressed exactly as Eq.(\ref{a31})]. 

Substituting Eqs.(\ref{e13}) and (\ref{a31}) into Eq.(\ref{e11}),
one derives the equations for $\varphi(\stackrel{\rightharpoonup 
}{q})$ 
and $f(\stackrel{\rightharpoonup }{q})$ in the meson rest 
frame\cite{t9}:
\begin{eqnarray}
\label{a26}
M\varphi _1(\stackrel{\rightharpoonup }{q})
&=&(E_1+E_2)\varphi _1(\stackrel{\rightharpoonup }{q})\nonumber 
\\
&&-\frac{E_1E_2+m_1m_2+\stackrel{%
\rightharpoonup }{q}^2}{4E_1E_2}\int 
d^3k(G_S(\stackrel{\rightharpoonup }{q}
, \stackrel{\rightharpoonup }{k})-4G_V(\stackrel{\rightharpoonup 
}{q}, %
\stackrel{\rightharpoonup }{k}))\varphi _1(\stackrel{\rightharpoonup 
}{k})\nonumber \\  
&&-\frac{(E_1m_2+E_2m_1)}{4E_1E_2}\int 
d^3k(G_S(\stackrel{\rightharpoonup }{q}, %
\stackrel{\rightharpoonup }{k})+2G_V(\stackrel{\rightharpoonup 
}{q}, %
\stackrel{\rightharpoonup }{k}))
\frac{m_1+m_2}{E_1+E_2}\varphi _1(\stackrel{\rightharpoonup 
}{k})\nonumber \\  
&&+\frac{E_1+E_2}{4E_1E_2} 
\int d^3kG_S(%
\stackrel{\rightharpoonup }{q}, \stackrel{\rightharpoonup 
}{k})(\stackrel{\rightharpoonup }{q}\cdot \stackrel{%
\rightharpoonup }{k})\frac{m_1+m_2}{E_1m_2+E_2m_1}
\varphi _1(\stackrel{\rightharpoonup }{k})\nonumber \\  
&&+\frac{%
m_1-m_2}{4E_1E_2} \int 
d^3k(G_S(\stackrel{%
\rightharpoonup }{q}, \stackrel{\rightharpoonup 
}{k})+2G_V(\stackrel{%
\rightharpoonup }{q}, \stackrel{\rightharpoonup }{k}))
(\stackrel{\rightharpoonup }{q}\cdot\stackrel{%
\rightharpoonup }{k})\frac{E_1-E_2}{E_1m_2+E_2m_1}\varphi 
_1(\stackrel{\rightharpoonup }{k}),~~~
\end{eqnarray}
where
\begin{equation}
\label{a27}
\varphi _1(\stackrel{\rightharpoonup 
}{q})=\frac{(m_1+m_2+E_1+E_2)(E_1m_2+E_2m_1)}{4E_1E_2(m
_1+m_2)}%
\varphi (\stackrel{\rightharpoonup }{q}),
\end{equation}
\begin{eqnarray}
\label{a28}
Mf_1(\stackrel{\rightharpoonup 
}{q})&=&(E_1+E_2)f_1(\stackrel{\rightharpoonup }{q})\nonumber 
\\
&&-\frac 1{4E_1E_2}\int d^3k(G_S(
\stackrel{\rightharpoonup }{q}, \stackrel{\rightharpoonup }{k})-
2G_V(%
\stackrel{\rightharpoonup }{q}, \stackrel{\rightharpoonup }{k}%
))(E_1m_2+E_2m_1)f_1(\stackrel{\rightharpoonup }{k})\nonumber \\  
&&-\frac{E_1+E_2}{4E_1E_2}\int 
d^3kG_S(\stackrel{\rightharpoonup }{q}, \stackrel{%
\rightharpoonup 
}{k})\frac{E_1m_2+E_2m_1}{E_1+E_2}
f_1(\stackrel{\rightharpoonup }{k})\nonumber \\  
&&+\frac{E_1E_2-m_1m_2+\stackrel{\rightharpoonup 
}{q}^2}{4E_1E_2\stackrel{%
\rightharpoonup }{q}^2} \int 
d^3k(G_S(%
\stackrel{\rightharpoonup }{q}, \stackrel{\rightharpoonup 
}{k})+4G_V(%
\stackrel{\rightharpoonup }{q}, \stackrel{\rightharpoonup 
}{k}))
(\stackrel{\rightharpoonup }{q}\cdot\stackrel{%
\rightharpoonup }{k})f_1(\stackrel{\rightharpoonup }{k})\nonumber \\  
&&-\frac{E_1m_2-E_2m_1}{4E_1E_2\stackrel{\rightharpoonup 
}{q}^2} \int d^3k(G_S(\stackrel{\rightharpoonup 
}{q}, %
\stackrel{\rightharpoonup }{k})-2G_V(\stackrel{\rightharpoonup 
}{q}, %
\stackrel{\rightharpoonup }{k}))
(\stackrel{\rightharpoonup }{q}\cdot\stackrel{\rightharpoonup 
}{k})\frac{E_1-E_2}{%
m_2+m_1}f_1(\stackrel{\rightharpoonup }{k})\nonumber \\ 
&&-\frac{E_1+E_2-m_2-m_1}{2E_1E_2\stackrel{\rightharpoonup 
}{q}^2} \int 
d^3kG_S(\stackrel{%
\rightharpoonup }{q}, \stackrel{\rightharpoonup }{k})
(\stackrel{\rightharpoonup }{q}\cdot\stackrel{\rightharpoonup }{k})^2
\frac
1{E_1+E_2+m_1+m_2}f_1(\stackrel{\rightharpoonup 
}{k})\nonumber \\ 
&&-\frac{m_2+m_1}{E_1E_2\stackrel{\rightharpoonup 
}{q}^2} \int
d^3kG_V(\stackrel{\rightharpoonup }{q}, \stackrel{\rightharpoonup 
}{k})%
(\stackrel{\rightharpoonup }{q}\cdot\stackrel{\rightharpoonup }{k})^2
\frac
1{E_1+E_2+m_1+m_2}f_1(\stackrel{\rightharpoonup }{k}),
\end{eqnarray}
where
\begin{equation}
\label{a29}
f_1(\stackrel{\rightharpoonup }{q})=-
\frac{m_1+m_2+E_1+E_2}{4E_1E_2}
f(\stackrel{\rightharpoonup }{q}).  
\end{equation}
Eqs.(\ref{a26}) and (\ref{a28}) can also be formally expressed as:

\begin{eqnarray}
\label{e22}
&&
(M-E_1-E_2)\varphi_{1}({\vec q})=\int d^{3} k\sum\limits_{{i=S},{V}}
F_{i}^{0^-} ({\vec q},{\vec k})G_i({\vec q},{\vec k})\varphi_{1}({\vec k}),
\nonumber\\
&&(M-E_1-E_2)f_{1}({\vec q})=\int d^{3} k\sum\limits_{{i=S},{V}}
F_{i}^{1^-} ({\vec q},{\vec k})G_i({\vec q},{\vec k})f_{1}({\vec k}).
\end{eqnarray}
In most cases, the interaction kernel is of the convolution type, i.e., 
$G(\vec{q},\vec{k})=G(\vec{q}-\vec{k})=G(\vec{p})$, where $\vec{p}
=\vec{q}-\vec{k}$ is the momentum exchanged between the quark and antiquark.
In the nonrelativistic limit for both quark and antiquark, 
Eq.~(\ref{a26}) and (\ref{a28}) can be expanded in terms
of ${\stackrel{\rightharpoonup }{q}}^2/{m_1}^2$ 
and ${\stackrel{\rightharpoonup }{q}}^2/{m_2}^2$,
and they are identical with the Schr\"odinger equation to
the zeroth order, and with the Breit equation to the first order. 

\section{Interaction kernel and retardation for confinement}

{\hskip 7mm}To solve Eq.(\ref{e3}) one must have a good command of the 
potential between 
two quarks. At present, the reliable information about the 
potential only comes from the 
lattice QCD result, which shows that the potential for a 
heavy quark-antiquark
pair $Q\bar{Q}$ in the static limit is well described by a long-ranged 
linear confining
potential ( Lorentz scalar $V_S$ ) and a short-ranged one gluon 
exchange potential
( Lorentz vector $V_V$ ), i.e,\cite{la,lb},
\begin{equation}
\label{a7}
{V_S(\stackrel{\rightharpoonup }{r})}=\lambda r ,
~~~~{V_V(\stackrel{\rightharpoonup }{r})}=-\frac 43{\alpha_s(r)
\over r}.
\end{equation}
The lattice QCD result for the $Q\bar{Q}$ potential is
supported by the
heavy quarkonium spectroscopy including both spin-independent and 
spin-dependent effects\cite{s1,s2,ding}.
Here, as the first step, we will employ the static potential below regardless of 
whether
the quarks are heavy or not
\begin{eqnarray}
\label{a9}
&&{V(r)}={V_S(r)+\gamma_{\mu}\otimes\gamma^{\mu} V_V(r)},\nonumber \\
&&{V_S(r)}={\lambda r\frac {(1-e^{-\alpha r})}{\alpha 
r}},\nonumber \\
&&{V_V(r)}=-{\frac 43}{\frac {\alpha_{s}(r)} r}e^{-\alpha r},
\end{eqnarray}
where the introduction of the factor $e^{-\alpha r}$ is to avoid 
the infrared(IR) divergence and also to incorporate 
the color screening 
effects of the dynamical light quark pairs on the ``quenched'' $Q\bar{Q}$ 
potential
\cite{t13}. It is clear that when $\alpha r\ll 1$ 
the potentials given in 
(\ref{a9}) become identical with that given in (\ref{a7}).
In momentum space the potentials are
\vskip 0.2cm
\begin{eqnarray}
\label{a12}
&&G( \stackrel{\rightharpoonup }{p})=G_S( \stackrel{%
\rightharpoonup }{p}) +\gamma_{\mu}\otimes \gamma^{\mu}
G_V( \stackrel{\rightharpoonup 
}{p}),\nonumber \\ 
&&G_S( \stackrel{\rightharpoonup }{p})=-\frac \lambda \alpha
\delta ^3( \stackrel{\rightharpoonup }{p})+\frac \lambda {\pi
^2}\frac 1{( \stackrel{\rightharpoonup }{p}^2+\alpha ^2) 
^2},\nonumber \\
&&G_V( \stackrel{\rightharpoonup }{p})=-\frac 2{3\pi^2}
\frac {\alpha_{s}(\stackrel{\rightharpoonup 
}{p})}{\stackrel{\rightharpoonup }{p}^2+\alpha ^2},
\end{eqnarray}
where $\alpha_{s}(\stackrel{\rightharpoonup }{p})$ is the well known 
running 
coupling constant and is assumed to become a constant of $O(1)$ as 
${\stackrel{\rightharpoonup }{p}}^2\rightarrow 0$
\begin{equation}
\label{a15}
\alpha _s( \stackrel{\rightharpoonup }{p}) =\frac{12\pi }{27}%
\frac 1{\ln ( a+\frac{\stackrel{\rightharpoonup }{p}^2}{\Lambda 
_{QCD}^2%
}) }.
\end{equation}
\vskip 0.2cm
The constants $\lambda$, $\alpha$, $a$, and $\Lambda_{QCD}$ are 
the parameters 
that characterize the potential.

Next, an important step is to take the retardation effect of scalar 
confinement into consideration. As disscussed in section I, the retardation 
effect of 
confinement will be approximatly treated by adding a retardation term  
$\frac{2 p_0^2}{\vec{p}^6}$ to the instantaneous part 
$\frac{1}{(\vec{p}^2)^2} $
as given in Eq.(\ref{3}),and $p_0^2$ will be treated to 
take its on-shell values which are obtained by assuming that the quarks are 
on their mass shells. Then this retardation term will become instantaneous
(but not convoluted). This modified scalar confinement  
potential will include the retardation
effect and become
\vskip 0.2cm
\begin{eqnarray}
\label{p24}
G_S(\vec{p})\rightarrow G_{S}(\vec{p},\vec{k})&=&
-\frac{\lambda}{\alpha}\delta^3(\vec{p}) +
\frac{\lambda}{\pi^2}\frac{1}{(\vec{p}^2 + \alpha^2)^2} \nonumber \\
&+& \frac{2 \lambda}{\pi^2}\frac{1}{(\vec{p}^2 + \alpha^2)^3}
\big(\sqrt{(\vec{p}+\vec{k})^2+m^2} - \sqrt{\vec{k}^2+m^2})^2 .
\end{eqnarray}
\vskip 0.2cm
This shows that the  retardation effect of confinement is 
taken into consideration  in the way that the interaction kernel depends 
not only on $\vec{p}$ (the momentum exchanged between quark and antiquark)
but also on $\vec{k}$ (the momentum of the quark itself). By
calculation below one may see that the retardation
effect is not negligible for 
$c\bar{c}$ states and become very significant for 
light-quark systems and we find it might be an 
useful remedy for the "intrinsic flaw" of the reduced BS equation 
with static scalar confinement.
In the computation in the next section we will use
\vskip 0.2cm
\begin{equation}
\label{a16}
{\lambda=0.183 GeV^{2}, ~~\alpha=0.06 
GeV,~~a=e=2.7183,~~\Lambda_{QCD}=0.15 GeV}, 
\end{equation}
\vskip 0.2cm
All these numbers are in the scopes of customarily usage.

\section{Results and discussions}

{\hskip 7mm}Based on the formulism above, we have calculated the mass 
spectrum of quarkonium including both heavy-
and light-quark systems. The numerical results with retardation are 
listed in Table I. The quark masses for the fit in Table I are
\begin{eqnarray}
\label{e24}
&&m_u=0.35GeV,~~  m_d=0.35GeV,~~ m_s=0.5GeV,\nonumber \\
&&m_c=1.65GeV,~~  m_b=4.83GeV .  
\end{eqnarray}
\vskip 1cm
\begin{center}
\tabcolsep 4mm
\begin{tabular}{|c|c|c|c|c|c|c|}\hline
 \multicolumn{7}{|c|}{$ 0^- $ meson masses}\\ \hline            
  \multicolumn{1}{|c|}{}&\multicolumn{2}{|c|}{$ 1S $} & \multicolumn{2}{|c|}{$ 2S $} & \multicolumn{2}{|c|}{$ 3S $ }\\ \hline
  \multicolumn{1}{|c|}{}& Fit& Data&Fit&Data&Fit&Data\\
  States & {\small (MeV)} &{\small (MeV)} &{\small (MeV)} & {\small (MeV)} & {\small (MeV)} & {\small (MeV)}\\ \hline
   $  u\bar{u},~ d\bar{d} $ & 500 &$ \pi (140)$ & 1252 &$\pi (1300)$  & 1611 &\\ 
    $ s\bar{s} $ &789 & &1559 & &1933 & \\ 
    $ c\bar{c} $ & 2976 &$\eta_c (2980) $ &  3657 & & 4032 & \\ 
    $ b\bar{b} $ & 9400 & & 9997 & & 10345 & \\ \hline
\multicolumn{7}{|c|}{$ 1^- $ meson masses}\\ \hline            
  \multicolumn{1}{|c|}{}&\multicolumn{2}{|c|}{$ 1S $} & \multicolumn{2}{|c|}{$ 2S $} & \multicolumn{2}{|c|}{$ 3S $}\\ \hline
  \multicolumn{1}{|c|}{}& Fit & Data& Fit& Data&Fit& Data\\
  States &  {\small (MeV)} & {\small (MeV)} & {\small (MeV)} & {\small (MeV)} & {\small (MeV)} & {\small (MeV)}\\ \hline
    $  u\bar{u} ,~ d\bar{d} $ & 763 &$\omega (782)$ & 1359 &$\omega (1420)$ & 1673 &$\omega (1662)$ \\ 
    $ s\bar{s} $ & 1025 &$\phi (1020)$ & 1649 &$\phi (1680)$ & 1989 &  \\ 
    $ c\bar{c} $ & 3119 &$J/\psi (3097)$  & 3701 &$\psi (3686)$ & 4062 &$\psi (4040)$ \\ 
    $ b\bar{b} $ & 9460 &$\Upsilon (9460)$ & 10013 &$\Upsilon (10023)$ & 10353 &$\Upsilon (10355)$ \\ \hline
\end{tabular}
\begin{center}
  {\small {\bf TABLE I.} Calculated mass spectra of $ b\bar{b}, c\bar{c}, 
  s\bar{s}, $ and $ u\bar{u} $ or $ d\bar{d} $  states 
using reduced Salpeter equation with retardation for scalar confinement. 
The Experiment data are taken from Ref.\cite{s9}.}  
\end{center}
\end{center}
\vskip 0.7cm
\begin{center}
\tabcolsep 4mm
\begin{tabular}{|c|c|c|c|}\hline
\multicolumn{1}{|c|}{}&\multicolumn{1}{|c|}{}& BS & BS \\
Level Spacings & Data & With~Retardation  & Without~Retardation \\
 \multicolumn{1}{|c|}{}& {\small (MeV)}& {\small (MeV)} & {\small (MeV)} \\ \hline
\multicolumn{4}{|c|}{ $ u\bar{u}~~d\bar{d} $~~states}\\ \hline
$\omega (2S)-\omega (1S)$&638 &596  &468  \\  
$\omega (3S)-\omega (1S)$&880 &910  &727  \\ \hline
\multicolumn{4}{|c|}{ $ s\bar{s} $ ~~states}\\ \hline
$\phi (2S)-\phi (1S) $&660 &624  &494  \\ 
$\phi (3S)-\phi (1S) $&?   &964  &782   \\ \hline
\multicolumn{4}{|c|}{ $ c\bar{c} $~~states}\\ \hline
$\psi (2S)-J/\psi$&589 &582  &522   \\ 
$\psi (3S)-J/\psi$&943 &943 &867  \\ \hline 
\multicolumn{4}{|c|}{ $ b\bar{b} $~~states}\\ \hline
$\Upsilon (2S)-\Upsilon (1S)$&563 &553  &536  \\ 
$\Upsilon (3S)-\Upsilon (1S)$&895 &893  &874  \\ \hline
\end{tabular}
\end{center} 
\begin{center}
{\small {\bf TABLE II.}~~The $2S-1S$ and $3S-1S$ energy level spacings of 
vector mesons with retardation and without retardation for scalar confinement}
\end{center}
\vskip 0.2cm
\par
For comparison with the results obtained without retardation, in Table II
we give a list of $2S-1S$ and $3S-1S$ energy level spacings for vector mesons 
in two cases, i.e., with retardation (using(\ref{p24})) and without retardation 
(using (\ref{a12})) for the scalar confinement potential.

From Table I and Table II we can clearly see the following.

$(1)$ The calculated level spacings without retardation are generally smaller 
than their experimental values. This trend is already appreciable for 
charmonium and becomes a serious problem for light quarkonium states. 
This result agrees with that obtained in Ref.\cite{s3}. We might
improve the fit by readjusting the parameters (e.g., by enlarging the 
string tension), and this may work for low-lying heavy quarkonium states
(e.g., $c\bar{c}$ states) but can not give a good global fit for high-lying 
states especially for light quarkonium states.

$(2)$ By adding the retardation term to scalar confinement potential the
calculated level spacings are significantly improved. The fit for $b\bar{b}$
and $c\bar{c}$ states is very good, and the fit for light vector mesons is 
also good, while the fit for light pseudoscalar mesons such as the pion is 
poor, which is probably due to the fact that the light pseudoscalar mesons
are essentially Goldstone bosons and therefore the instantaneous and 
on-shell approximations no longer work well for them.

As emphasized in Section I, the approximate treatment for the retardation 
effect (in particular, the on-shell approximation) of scalar confinement 
should be good for heavy quarkonium states. Indeed, it has been shown\cite{ch}
that for heavy quarkonium in nonrelativistic expansion the role of the 
retardation is just to cancel the troublesome term, 
$-\frac{1}{2 m^2}(\vec{p}^2
S + S\vec{p}^2)$ in Eq.(\ref{1}) and then remove the bad disturbance to the mass
spectrum. 

For light quarkonium states with constituent quark masses $m_u=m_d\approx 
350 MeV$,the retardation effect bocomes even more significant. In these systems 
nonrelativistic expansion is no longer good, but we can see the 
physical effect of retardation through an extreme case. That is the zero 
quark mass limit, which has been used for analysing the "intrinsic flaw"
of scalar confinement, indicating that light quarks can only have very weak 
confinement if it is an instantaneous Lorentz scalar potential
\cite{s3}. To see how the "retardation" part changes the trend of light
quarks seeing a weaker confining potential than heavy quarks at large 
distances, it is useful to consider again the limit of zero quark mass. As
light quark systems are more sensitive than heavy quark systems to the 
behavior of interaction at large distance, we will restrict our discussion
only on the scalar confinement potential part.

In the zero-quark mass limit, as $m_{q}\rightarrow 0$, the coefficients for the 
scalar potential $G_s$ in Eq.(\ref{e22}) for the $0^-$ and $1^-$
mesons will reduce to 
\begin{eqnarray}
\label{bb1}
&&F_{S}^{0^-}(\vec{q},\vec{k})\longrightarrow-\frac{1}{2}(1-\frac{\vec{q}
\cdot\vec{k}}{q k}),\nonumber\\
&&F_{S}^{1^-}(\vec{q},\vec{k})\longrightarrow -\frac{1}{2}
\frac{(\vec{q}\cdot\vec{k})}{q^2}(1-\frac{\vec{q}
\cdot\vec{k}}{q k}).
\end{eqnarray}
where $q=|\vec{q}|$, $k=|\vec{k}|$. It is clear that these coefficients will
vanish when $\vec{q}\rightarrow\vec{k}$. On the other hand, however, the 
static linear confining potential in momentum space behaves as $G_s(\vec{q}-
\vec{k})\propto(\vec{q}-\vec{k})^{-4}$ and is strongly weighted as $\vec{q}
\rightarrow \vec{k} $ in Eq.(\ref{e22}). This is the reason why 
the light quarks can only have weak confinement, which leads to very narrow 
energy level spacings for light quarkonium states. This bad situation will be
changed if the retardation is taken into account. In fact, the covariant form
of confinement interaction may take the form $G_{S}( q , k ) \propto[(\vec{q}
-\vec{k})^2 - (q_0 - k_0)^2]^{-2}$ and in the on-shell approximation that
$q_0^2=m^2 + \vec{q}^2,~k_0^2=m^2 + \vec{k}^2$ it becomes

\begin{equation}
G_{S}( q , k ) \propto[-2 m^2 + m^2\frac{k}{q} +m^2\frac{q}{k}  +
2 (qk - \vec{q}\cdot\vec{k})]^{-2} ,
\end{equation}
where in the zero-quark mass limit $p, k \gg m\rightarrow 0$. We can see
immediately that it has two distinct features from the static confining 
potential $G_{S}(\vec{q},\vec{k})\propto [(\vec{q}-\vec{k})^2]^{-2}$. First, 
with retardation the scalar interaction $G_{S}(q, k)$ is strongly weighted 
when $\vec{q}$ and $\vec{k}$ are co-linear ($\vec{q}\parallel\vec{k}$) that
leads to $\vec{q}\cdot\vec{k}=qk$, whereas the static linear potential only 
peaks at $\vec{q}=\vec{k}$. This indicates that the former is strongly 
weighted in a much wider kinematic region than the latter. Second, when 
$\vec{q}\parallel\vec{k}$, $G_S(q, k)\propto O(m^{-4})$, this mass dependence,
which is absent in the static linear potential, will enhance confinement 
interaction and overwhelm the suppression
factor apppearing in the coeficients $F_{S}^{0^-}$ and $F_{S}^{1^-}$. As a 
result, even in the zero-quark-mass limit the effective scalar interaction 
will not be weakened, and this is just due to the retardation effect in the 
scalar interaction.

In practice, for the constituent quark model, which is essentially used in 
the present work, the quark mass can not be zero, and the on-shell approximation 
may not be as good as for heavy quark systems. But in any case the analysis given 
above in the zero mass limit is useful for understanding the qualitative 
feature of the retardation effect in quark confinement.

In this paper, we have tried to clarify the problem pointed out by Durand 
{\it et al.} for the static scalar confinement in reduced Salpeter equation. 
The "intrinsic flaw" of the Salpeter equation with static scalar confinement 
could  be remedied to some extent by taking the retardation effect of the 
confinement into consideration. In the on-shell approximation for the 
retardation term of linear confinement, the notorious trend of narrow level 
spacings for quarkonium states especially for light quarkonium states is found
to be removed. A good fit for mass spectrum  of S-wave heavy and light 
quarkonium states (except the light pseudoscalar mesons) is obtained using 
one-gluon exchange potential and the scalar linear confinement potential with 
retardation  taken into account. Although for light quark systems the on-shell
appproximation may not be good, the qualitative feature of the retardation 
effect is still manifest.  We may then conclude that at phenomenological
level including the retardation effect into the scalar confinement may be 
necessary and significant. Nevertheless, it is still premature to assess 
whether or not quark confinement is really represented by the scalar 
exchange of the form of $(\vec{p}^2-p_0^2)^{-2}$, as suggested by some 
authors as the dressed gluon propogator to implement quark confinement.
We hope that our investigation can provide some useful information for the 
understanding of confinement. Further discussions concening heavy-light 
mesons will be given in another publication.
\vskip 1cm
\begin{center}
\bf\large\bf{Acknowlegement}
\end{center}

This work was supported in part by the National Natural Science Foundation
of China, the State Education Commission of China and the State Commission
of Science and Technology of China.

%\vskip 0.5 in

\end{document}